\documentclass[a4paper,twoside]{article}
\usepackage{amsmath,amssymb,amsthm}
\usepackage{pstricks}
\usepackage{graphicx}

\newcommand{\ket}[1]{| #1 \rangle}
\newcommand{\bra}[1]{\langle #1 |}

\def\shuff#1#2{\mathbin{
\hbox{\vbox{ \hbox{\vrule \hskip#2 \vrule height#1 width 0pt
}%
\hrule}%
\vbox{ \hbox{\vrule \hskip#2 \vrule height#1 width 0pt
\vrule }%
\hrule}%
}}}

\def\shuf{{\mathchoice{\shuff{7pt}{3.5pt}}%
{\shuff{6pt}{3pt}}%
{\shuff{4pt}{2pt}}%
{\shuff{3pt}{1.5pt}}}}%
\def\shuffle{\,\shuf\,}

\newcommand{\Q}{\mathbb{Q}}

\newcommand{\N}{\mathbb{N}}

\newtheorem{prop}{Proposition}[section]
\newtheorem{rem}[prop]{Remark}
\newtheorem{lem}[prop]{Lemma}
\newtheorem{cor}[prop]{Corollary}
\newtheorem{thm}{Theorem}[section]
\newtheorem{de}[prop]{Definition}

\newcommand{\Hint}{H_{\mathrm{I}\epsilon}}
\newcommand{\Heff}{H_{\mathrm{eff}}}
\newcommand{\PsiGL}{\Psi_{\mathrm{GL}}}

\newcommand{\ee}{{\mathrm{e}}}
\newcommand{\dd}{{\mathrm{d}}}

\newcommand{\bs}{{\bar{\sigma}}}
\newcommand{\bbe}{{\bar{\beta}}}

\newcommand{\bt}{{\bar{\tau}}}

\begin{document}
\title{\textbf{Hyperoctahedral Chen calculus for effective Hamiltonians}}

\author{Christian Brouder\\
 Institut de Min\'eralogie et de Physique des Milieux
  Condens\'es, \\CNRS UMR7590,
 Universit\'es Paris 6 et 7, IPGP,\\ 140 rue de Lourmel,
  75015 Paris, France.\\\\
Fr\'ed\'eric Patras\\
Laboratoire J.-A. Dieudonn\'e, CNRS UMR 6621,\\
Universit\'e de Nice,\\
Parc Valrose, 06108 Nice Cedex 02, France.
}%
\maketitle
\begin{abstract}
We tackle the problem of unraveling the algebraic structure of computations of effective Hamiltonians. This is an important subject in view of applications to chemistry, solid state physics or quantum field theory.
We show, among other things, that the correct framework for these computations is provided by the hyperoctahedral group algebras. We define several structures on these algebras and give various applications. For example, we show that the adiabatic evolution operator (in the time-dependent interaction representation of an effective Hamiltonian) can be written naturally as a Picard-type series and has a natural exponential expansion.
\end{abstract}

\section*{Introduction}

We start with a short overview of the classical theory of Chen calculus,
that is, iterated integral computations. The subject is classical but is
rarely presented from the suitable theoretical perspective -that is, emphasizing the role of the convolution product on the direct sum of the symmetric groups group algebras. We give therefore a brief account of the theory that takes into account this point of view -this will be useful later in the article. Then, we recall the construction of effective Hamiltonians in the time-dependent interaction representation.

The third section is devoted to the investigation of the structure of
the hyperoctahedral group algebras. Although we are really interested in
the applications of these objects to the study of effective
Hamiltonians, and although the definitions we introduce are motivated by
the behavior of the iterated integrals showing up in this setting, we
postpone the description of the way the two theories interact to a later
stage of the article. Roughly stated, we show that the descent algebra
approach to Lie calculus, as emphasized in Reutenauer's
\cite{reutenauer1993} can be lifted to the hyperoctahedral setting. This
extends previous works \cite{MnR95,ABN04,BH06,ANT08,NT08} on the subject
and shows that these results (focussing largely on Solomon's algebras of hyperoctahedral groups and other wreath product group algebras) are naturally connected to the study of physical systems through the properties of their Hamiltonians and of the corresponding differential equations, very much as the classical theory of free Lie algebras relates naturally to the study of differential equations and topological groups. Notice however that the statistics we introduce here on hyperoctahedral groups seems to be new --and is different from the statistics naturally associated to the noncommutative representation theoretic approach to hyperoctahedral groups, as it appears in these works.

The fourth section studies the effective adiabatic evolution operator and shows that it can be expanded as a generalized Picard series by means of the statistics introduced on hyperoctahedral groups\footnote{Picard series are often referred to as Dyson or Dyson-Chen series in the literature, especially in contemporary physics, but we prefer to stick to the most classical terminology}. As a corollary, we derive in the last section an exponential expansion for the evolution operator. Such expansions are particularly useful in view of numerical computations, since they usually lead to approximating series converging much faster than the ones obtained from the Picard series.

Acknowledgements. We thank Kurusch Ebrahimi-Fard, Jean-Yves Thibon and an anonymous referee. Their comments and suggestions resulted into several improvements with respect to a preliminary version of this article.

\section{The algebra of iterated integrals}\label{sect1}

Let us recall the basis of Chen's iterated integrals calculus, starting with a first order linear differential equation (with, say, operator or matrix coefficients):
$$A'(t)=H(t)A(t), \ A(0)=1$$
The solution can be expanded as the Picard series:
$$A(t)=1+\int\limits_0^tH(x)dx+\int\limits_0^{t}\int\limits_0^{t_1}H(t_1)H(t_2)dt_1dt_2+...+\int\limits_{\Delta_n^t}  H(t_1)...H(t_n)+...$$
where $\Delta_n^t:=\{0\leq t_n\leq ...\leq t_1\leq t\}$
and where the measure $d t_1\dots d t_n$ is implicit.
Notice, for further use, that when we allow for a general initial condition $A(x)=1$, with possibly $x=-\infty$, the integration simplex $\Delta_n^t$ should be replaced by $\Delta_n^{[x,t]}:=\{x\leq t_n\leq ...\leq t_1\leq t\}.$

Solving for $A(t)=\exp(\Omega(t))$ (see \cite{BMP69,MP70}), and more generally any computation with $A(t)$, requires the computation of products of iterated integrals of the form:
$$<\sigma> :=\int\limits_{\Delta_n^t}  H(t_{\sigma (1)})...H(t_{\sigma
(n)}),\, \sigma =(\sigma(1),...,\sigma(n))\in S_n,$$
where $S_n$ stands for the symmetric group of order $n$. Notice that we represent an element $\sigma$ in $S_n$ by the sequence $(\sigma(1),...,\sigma(n))$.

In general, for any $\mu=\sum\limits_n\sum\limits_{\sigma\in S_n}\mu_\sigma\cdot \sigma\in {\bf S}:=\bigoplus\limits_n\Q[S_n]$, the direct sum of the group algebras of the symmetric groups $S_n$ over the rationals, we will write $<\mu>$ for $\sum\limits_n\sum\limits_{\sigma\in S_n}\mu_\sigma\cdot <\sigma>$. This allows, for example, to write $A(t)$ as $<I>$, where $I:=\sum\limits_n(1,...,n)$ is the formal sum of the identity elements in the symmetric group algebras.

The formula for the product of $<\sigma>$ with $<\beta>$ is a variant of Chen's formula for the product of two iterated integrals of functions or of differential forms (a proof of the formula will be given in Section~\ref{sect3} in a more general framework; the formula also holds for arbitrary integration bounds, that is when $\Delta_n^t$ is replaced by $\Delta_n^{[x,t]}$):
$$<\sigma>\cdot <\beta> =<{\sigma\ast \beta}>,$$
where $\sigma\ast\beta$ is the convolution product\footnote{This is one possible definition of the convolution product, there are several equivalent ones that can be obtained using the various natural set automorphisms of the symmetric groups (such as inversion or conjugacy by the element of maximal length). They result into various (but essentially equivalent) associative algebra structures on the direct sum of the symmetric groups group algebras, see e.g. \cite{MR95}} of the two permutations, that is, for $\sigma\in S_n,\ \beta\in S_m$:
$\sigma\ast\beta$ is the sum of the $\binom{n+m}{n}$
permutations $\gamma\in S_{n+m}$ with
$st(\gamma(1),...,\gamma(n))=(\sigma(1),...,\sigma(n))$ and
$st(\gamma(n+1),...,\gamma(n+m))=(\beta(1),...,\beta(m))$. Here, $st$
stands for the standardization map, the action of which on sequences is
obtained by replacing $(i_1,...,i_n), i_j\in\N^\ast$ by the (necessarily
unique) permutation $\sigma\in S_n$, such that $\sigma(p)<\sigma(q)$ for $p<q$ if and only if $i_p\leq i_q$. 
In words, each number $i_j$ is replaced by the position of $i_j$
in the increasing ordering of $i_1$, \dots, $i_n$.
If we take the example of $(5,8,2)$, the position of $5$, $8$ and $2$
in the ordering $2 < 5 < 8$ is $2$, $3$ and $1$. Thus, $st(5,8,2)=(2,3,1)$.
For instance,
\begin{eqnarray*}
(2,3,1)\ast (1) &=& (2,3,1,4)+(2,4,1,3)+(3,4,1,2)+(3,4,2,1),\\
(1,2)\ast (2,1) &=& (1,2,4,3)+(1,3,4,2)+(1,4,3,2)+(2,3,4,1)+(2,4,3,1)
+(3,4,2,1).
\end{eqnarray*}

The convolution product relates to the shuffle product $\shuffle$ \cite{reutenauer1993} as it appears in Chen's work \cite{chen1973}
and in the parametrization of the product of iterated integrals of functions or differential forms.
Indeed, for $\sigma , \beta$ as above:
$$\sigma^{-1}\ast\beta^{-1}=(\sigma\shuffle\beta [n])^{-1}$$
where $\beta [n]$ stands for the sequence $(\beta(1)+n,...,\beta(m)+n)$.
Recall, for completeness sake, that the shuffle product $A\shuffle B$ of two words (or sequences) $A=aA'=aa_2...a_k$, $B=bB'=bb_2...b_l$ is defined recursively by 
$$A\shuffle B=a (A'\shuffle B)+b(A\shuffle B')$$
we refer to \cite{sch58} and \cite{reutenauer1993} for details.
Associativity of $\ast$ follows immediately from the definition,
the unit is $1\in {\bf S}_0=\Q$, and the graduation on 
${\bf S}=\bigoplus\limits_n\Q[S_n]$ is compatible with $\ast$, so that:

\begin{lem}
The convolution product provides $\bf S$ with the structure of a graded connected associative (but noncommutative) unital algebra.
\end{lem}

For completeness, recall that connected means simply that ${\bf
S}_0=\Q$. 
In fact, $\bf S$ carries the richer structure of a Hopf algebra, and as such is referred to as the Malvenuto-Reutenauer Hopf algebra \cite{MR95}.
From the point of view of the theory of noncommutative symmetric functions, the elements of $\bf S$ should be understood as free quasisymmetric functions \cite{DHT02}.
This definition of the convolution product on $\bf S$ allows, for example, to express simply the coefficients of the continuous Baker-Campbell-Hausdorff formula (compare with the original solution \cite{MP70}):
$$\Omega(t)=<\log(I)>.$$
Here $\log(I)$ identifies, in $\bf S$, with the formal sum of Solomon's
(also called Eulerian or canonical) idempotents \cite{solomon1968}. We refer to
\cite{pat92,reutenauer1993,pat93,pat94,G95} for an explanation and a
Hopf algebraic approach to these idempotents and, more generally, for a
Hopf algebraic approach to Lie computations. We will return later with
more details to Solomon's idempotent but mention only, for the time being, that one of the main purposes of the present article is to extend these ideas to the more general framework required by the study of effective Hamiltonians.

\section{Iterated integrals in time-dependent perturbation theory}

The ultimate aim of quantum physics is the knowledge of the eigenvalues
and eigenstates of the Hamiltonian $H$ describing a physical system.
In most cases, the eigenvalue problem cannot be solved exactly, but
the eigenvalues and eigenstates of a simpler and closely related Hamiltonian $H_0$
are known, at least numerically. Then, $H$ can be rewritten $H=H_0+V$,
where $V$ is referred to as the perturbation term.
For example, in molecular physics,
$H_0$ is the Hamiltonian describing the interactions of $N_e$
electrons with $N_n$ nuclei, whereas $H$ describes
the interaction of the $N_e$ electrons with themselves 
\emph{and} with the $N_n$ nuclei.
In that case, the perturbation $V$ describes the electron-electron interaction, this simple approach paving the way
to most of the numerical methods in the field.

In other terms, perturbation theory provides a systematic way to calculate
an eigenstate of $H$ from an eigenstate of $H_0$.
In the time-dependent perturbation theory, we first define a
time-dependent Hamiltonian $H(t)=H_0 + \ee^{-\epsilon |t|} V$.
When $\epsilon$ is small, this means
physically that the interaction is very slowly switched on from
$t=-\infty$ where $H(-\infty)=H_0$ to $t=0$
where $H(0)=H$. 
The basic idea is that, if $\epsilon$
is small enough, then an eigenstate of $H_0$
can be transformed into an eigenstate of $H$
by the time-dependent perturbation $\ee^{-\epsilon |t|} V$.
For example, the ground state $|\Phi_0>$ (the eigenstate associated to the lowest eigenvalue $E_0$
of $H_0$) should hopefully be transformed into the ground state of $H$.
We also assume for the time being that the ground state is non degenerate, that is that  
the eigenspace associated to the highest eigenvalue $E_0$ is one-dimensional. We write from now on $|\Phi_i>$
for the eigenvectors of $H_0$ and assume that the eigenvalues $E_i$ are ordered increasingly ($E_i\leq E_{i+1}$).

To implement this picture of perturbation theory, the time-dependent
Schr\"odinger equation
$i\partial \ket{\Psi_S(t)}/\partial t=H(t) \ket{\Psi_S(t)}$
should be solved.
However, the solutions $\ket{\Psi_S(t)}$ of this equation
vary like $\ee^{-i E_0 t} \ket{\Phi_0}$ for large negative $t$
(where $\ket{\Phi_0}$ satisfies 
$H_0 \ket{\Phi_0} = E_0 \ket{\Phi_0}$) 
and have no limit for $t\to -\infty$.
To compensate for this time variation, 
one looks instead for $\ket{\Psi(t)}=\ee^{i H_0 t} \ket{\Psi_S(t)}$ that satisfies
$i\partial \ket{\Psi(t)}/\partial t=\Hint(t) \ket{\Psi(t)}$, with
$\Hint=\ee^{iH_0t} V \ee^{-iH_0t} \ee^{-\epsilon|t|}$.
Now $\Hint(-\infty)=0$, and $\ket{\Psi(-\infty)}$ makes sense.
Using $\Hint$, we can start consistently from the ground state $\ket{\Phi_0}$
of $H_0$ and solve the time-dependent 
Schr\"odinger equation with the boundary condition
$\ket{\Psi(-\infty)}=\ket{\Phi_0}$. When no eigenvalue crossing takes place,
$\ket{\Phi_0}$ should be transformed into the ground state
$\ket{\Psi(0)}$ of $H$.

At first sight, solving the time-dependent Schr{\"o}dinger equation for 
$\ket{\Psi(t)}$ does not look simpler than solving the
time-independent Schr{\"o}dinger equation with the Hamiltonian $H$.
However, if $V$ is small enough, fairly accurate approximations
of the true eigenstates can be obtained from the first terms of
the perturbative expansion of $\ket{\Psi(t)}$.
In general, instead of calculating directly $\ket{\Psi(t)}$, it is 
convenient to define the unitary operator
$U_\epsilon(t)$ as the solution of
$i\partial U_\epsilon(t)/\partial t=\Hint(t) U_\epsilon(t)$, with the boundary
condition $U_\epsilon(-\infty)=1$. Thus, 
$\ket{\Psi(t)}=U_\epsilon(t)\ket{\Phi_0}$.
The connection with iterated integrals appears when solving iteratively
the equation for $U_\epsilon(t)$:
\begin{eqnarray*}
U_\epsilon(t) &=& 1 + (-i) \int_{-\infty}^t \dd t_1 \Hint(t_1) +
 (-i)^2 \int_{-\infty}^t \dd t_1 \int_{-\infty}^{t_1} \dd t_2 \Hint(t_1) \Hint(t_2) + \dots
\end{eqnarray*}
A straightforward calculation~\cite{Goldstone} gives us
\begin{eqnarray}
U_\epsilon(0) \ket{\Phi_0} &=& \ket{\Phi_0} + \sum_{n=1}^\infty \sum_{i_1\dots i_n}
   \frac{\ket{\Phi_{i_n}}\bra{\Phi_{i_n}} V \ket{\Phi_{i_{n-1}}} 
   \dots
   \bra{\Phi_{i_2}} V \ket{\Phi_{i_{1}} }
   \bra{\Phi_{i_1}} V \ket{\Phi_{0}} }
   {(E_0-E_{i_n}+ni\epsilon)\dots (E_0-E_{i_2}+2i\epsilon)
   (E_0-E_{i_1}+i\epsilon) } ,
\label{Goldstone}
\end{eqnarray}
where we use the
completeness relation $1=\sum_i \ket{\Phi_{i}} \bra{\Phi_{i}}$.
However, it immediately appears that this expression has no limit for
$\epsilon\to 0$ because the sum over all $i_p$ contains
the terms $i_p=0$, for which the denominator has a factor $pi\epsilon$.
In 1951, Gell-Mann and Low \cite{GellMann} conjectured that
\begin{eqnarray*}
\ket{\PsiGL} &=& \lim_{\epsilon\to0}\frac{U_\epsilon(0)\ket{\Phi_0}}
   {\bra{\Phi_0} U_\epsilon(0)\ket{\Phi_0}}
\end{eqnarray*}
exists and is an eigenstate of $H$. The fact that the Gell-Mann and Low
formula indeed removes all the divergences of equation (\ref{Goldstone})
was proved much later by Nenciu and Rasche~\cite{Nenciu}. 

The above scheme works nicely when the ground state of $H_0$
is non degenerate. When it is degenerate, that is when the eigenspace $E_0$ 
associated to the lowest eigenvalue of $H_0$ has dimension $> 1$, the problem
is more subtle, see \cite{morita,bulaevskii,ms}. To understand the algebraic phenomena underlying 
time-dependent perturbation theory for degenerate systems is actually the purpose of the present article.

Let us write $P$ for the 
projection on this eigenspace. The natural extension of the Gell-Mann and Low 
formula then reads as a definition of a ``Gell-Mann and Low'' operator acting 
on the degenerate eigenspace $E_0$:
\begin{eqnarray}
U_{GL} &:=& \lim\limits_{\epsilon \to 0}{U_\epsilon(0)P}(PU_\epsilon(0)P)^{-1},
\label{UGL}
\end{eqnarray}
or, $U_{GL} = \lim\limits_{\epsilon \to 0}U_{GL}(\epsilon),\ U_{GL}(\epsilon):=
{U_\epsilon(0)P}(PU_\epsilon(0)P)^{-1}.$
It can be shown that the operator $PU_\epsilon(0)P$ is invertible within the image of $P$
if no vector in the image of $U_\epsilon(0)P$ is annihilated by $P$~\cite{Bloch}.
We will assume that this property is satisfied by the systems we consider.
The operator $U_{GL}$ was first proposed by Morita in 1963~\cite{morita}. It shows up e.g. 
in the time-dependent interaction representation of the effective Hamiltonian
$\Heff:=\lim\limits_{\epsilon \to 0}P H U_\epsilon(0)P[PU_\epsilon(0)P]^{-1}$ classically 
used to solve the eigenvalue problem.

This operator $U_{GL}$ is the one we will be interested in here, postponing to further work the analysis of 
concrete applications to the study of degenerate systems. 
In other terms,
we will investigate and unravel the fine algebraic structure of the iterative
expansions of $U_\epsilon(t)$ and $U_{GL}(\epsilon)$.

\section{Wreath product convolution algebras}\label{sect3}

Let us explain further our motivation. In  the previous section, we
observed that the study of effective Hamiltonians leads to the study of
Picard-type expansions involving the operators $\Hint(t)$ and
$P\Hint(t)$ or, equivalently, $A(t):=-i(1-P)\Hint(t)$ and
$B(t):=iP\Hint(t)$. Expanding these expressions will lead to the study
of iterated integrals involving the two operators $A(t)$ and $B(t)$ such
as, say: $\int\limits_{\Delta_3^t}A(t_2)B(t_3)A(t_1)$. The idea
underlying the forthcoming algebraic constructions is to encode such an
expression by a signed permutation and to lift computations with
iterated integrals to an abstract algebraic setting: in the previous
example, the signed permutation would be $(2,\bar{3},1)$ (see below for
precise definitions). 

In more abstract (but equivalent) terms, iterated integrals on two
operators are conveniently encoded by elements of the hyperoctahedral
groups, whereas iterated integrals on $k$ operators would be conveniently encoded by more general colored permutations or elements of the wreath product of $S_n$ with the cyclic group of order $k$. Recall the definition of the hyperoctahedral group $B_n$ of
order $n$. The hyperoctahedral group is the group defined either as the
wreath product of the symmetric group of order $n$ with the cyclic group
of order 2, or, in a more concrete way, as the group of ``signed permutations'' the elements of which are written as sequences of integers $i\in \N^\ast$ and of integers with an upper bar $\bar{i}, i\in\N^\ast$, so that, when the bars are erased, one recovers the expression of a permutation. The composition rule is the usual one for permutations, together with the sign rule for bars: for example, if $\bar\sigma\in B_3 =(2,\bar{3},1)$ and $\bar\beta =(\bar{3},1,\bar{2})$, then:
$$\bar\beta\circ\bar\sigma (2)=\bar\beta (\bar{3})=\bar{\bar{2}}=2,$$
$$\bar\beta\circ\bar\sigma (3)=\bar\beta (1)={\bar{3}}.$$

By analogy with $\bf S$, we equip ${\bf B}:=\bigoplus\limits_nB_n$ with the structure of a graded connected (associative but noncommutative) algebra with a unit. This algebra structure agrees with the ones introduced in \cite{ABN04,NT08,ANT08} (possibly up to an isomorphism: for example, the relationship between the product we consider and the shifted shuffle product of \cite{NT08} reflects the relationship between the convolution and (shifted) shuffle product on $\bf S$ that we recalled in the first section of the present article).

The standardization $st$ of a signed sequence $\bar{w}$ (i.e. a sequence of integers and of integers marked with an upper bar) is defined analogously to the classical standardization, except for the fact that upper bars are left unchanged (or, equivalently, have to be reintroduced at their initial positions after the standardization of the sequence $w$ has been performed, where we write $w$ for $\bar{w}$ where the upper bars have been erased). For example, $st(\bar 2,7,\bar 1,2)=(\bar 2,4,\bar 1,3)$.

\begin{de}
Let $\bar{\sigma},\bar{\beta}$ belong to $B_n$, resp. $B_m$. Their convolution product is defined by:
$$\bs\ast\bbe :=\sum\limits_\bt \bt$$
where $\bt$ runs over the $\binom{n+m}{n}$ 
elements of $B_{n+m}$ with 
$st(\bt(1),...,\bt(n))=\bs$, $st(\bt(n+1),...,\bt(n+m))=\bbe$.
\end{de}
For instance,
\begin{eqnarray*}
(\bar2,3,1)\ast (\bar1) &=& (\bar2,3,1,\bar4)+(\bar2,4,1,\bar3)+
(\bar3,4,1,\bar2)+(\bar3,4,2,\bar1),\\
(1,\bar2)\ast (2,\bar1) &=& (1,\bar2,4,\bar3)+(1,\bar3,4,\bar2)+
(1,\bar4,3,\bar2)+(2,\bar3,4,\bar1)+(2,\bar4,3,\bar1)
+(3,\bar4,2,\bar1).
\end{eqnarray*}
Notice that this definition is dictated, for us, by iterated integrals computations, similarly to the classical one-Hamiltonian case dealt with in the first section. Indeed,
let $A(t),B(t)$ be two time-dependent operators. For $\bar{\sigma}\in B_n$, let
us write $<{\bar\sigma}>$ for the iterated integrals obtained by the usual
process, with the extra prescription that upper indices (empty set or bar) in
$\bar{\sigma}$ indicate that the operator used at the corresponding level of the
integral is $A$ or $B$, so that e.g., $\bar{\sigma}=(\bar 3,1,\bar 2)$ is associated to: $\int\limits_{\Delta_3^t}B(t_3)A(t_1)B(t_2)$. For an arbitrary $\bar\gamma=\sum\limits_n\sum\limits_{\bar\sigma\in B_n}a_{\bar\sigma}\cdot\bar\sigma\in \bf B$, we write $<{\bar\gamma}>$ for $\sum\limits_n\sum\limits_{\bar\sigma\in B_n}a_{\bar\sigma}\cdot <{\bar\sigma}>$.

\begin{prop}
\label{HsHb}
The product of two iterated integrals $<{\bs}>\times <{\bbe}>$ is given by:
$$<{\bs}>\times <{\bbe}>=<{\bs\ast\bbe}>$$
\end{prop}

This formula is a noncommutative variant of the classical Chen formulas for the product of iterated integrals of differential forms \cite{chen1973}. It includes as a particular case the formula for the product of two iterated integrals depending on a single time-dependent Hamiltonian given in the first section of the article. We detail the proof for the sake of completeness, and since the formula is crucial for our purposes.

For a permutation $\bar\sigma$ we denote by $\sigma$ the same permutation 
without
bars (e.g. if $\bar\sigma=(\bar2,3,\bar1)$, then $\sigma=(2,3,1)$)
and we define
$X(t_{\sigma(i)})=A(t_{\sigma(i)})$ if $\bar\sigma(i)$ has no bar and
$X(t_{\sigma(i)})=B(t_{\sigma(i)})$ if $\bar\sigma(i)$ has a bar.
Therefore,

\begin{eqnarray*}
<{\bs}>\times <{\bbe}> &=&
\int_0^t dt_1 \dots \int_0^{t_{n-1}} dt_n X(t_{\sigma(1)})\dots X(t_{\sigma(n)})
\int_0^t dt_{n+1} \dots \int_0^{t_{n+m-1}} dt_{n+m}
 X(t_{n+\beta(1)})\dots X(t_{n+\beta(m)}).
\end{eqnarray*}
By Fubini's theorem, this can be rewritten as the integral of
$ X(t_{\sigma(1)})\dots X(t_{n+\beta(m)})$ over the domain
$\Delta^t_n\times\Delta^t_m$. The idea is now to rewrite this domain
as a sum of $\binom{n+m}{n}$ domains isomorphic to $\Delta^t_{n+m}$.
For instance, the product of the domain
$0\le t_n \le \dots \le t_1 \le t$ with the domain $0\le t_{n+1} \le t$
is the sum of the $n+1$ domains obtained by inserting
$t_{n+1}$ between $0$ and $t_1$, then between $t_1$ and $t_2$,
up to between $t_n$ and $t$. 
More generally the product of
$\Delta^t_n$ by $\Delta^t_m$ is the sum of all the domains
obtained by ``mixing'' the two conditions
$0\leq t_{n} \le \dots \le t_1 \le t$ and
$0\leq t_{n+m} \le \dots \le t_{n+1} \le t$,
i.e. by ordering the $n+m$ variables $t_i$ so that
these conditions are satisfied.
If $\rho(i)$ is the position of variable $t_i$ in 
one of these orderings (where the variables are ordered from the largest to the
smallest), the conditions imply that $\rho(1) < \dots < \rho(n)$
and $\rho(n+1) < \dots < \rho(n+m)$.
For example, if $0 \le t_2 \le t_1 \le t$ and
$0 \le t_4 \le t_3 \le t$, for the domain
$0 \le t_4 \le t_2 \le t_1 \le t_3 \le t$,
$t_3$ is in the first place (i.e. largest), $t_1$ in the second,
$t_2$ in the third and $t_4$ in the fourth (smallest), and the
permutation is $\rho=(2,3,1,4)$.
In general, we get:
$$\Delta^t_n\times \Delta^t_m=\bigcup\limits_{\tau}\{(t_{\tau(1)},...,t_{\tau(n+m)})|0\leq t_{n+m}\leq ...\leq t_1\leq t\},$$
where $\tau$ runs over the permutations in $S_{n+m}$ such that
$\tau(1) < \dots < \tau(n)$ and
$\tau(n+1) < \dots < \tau(n+m)$. 
Equivalently, $\tau $ runs over the permutations such that:
$st(\tau(1),\dots,\tau(n))=(1,\dots,n)$ and
$st(\tau(n+1),\dots,\tau(n+m))=(1,\dots,m)$. Now, 

$\int\limits_{\{(x_1=t_{\tau(1)},...,x_{n+m}=t_{\tau(n+m)})|0\leq t_{n+m}\leq ...\leq t_1\leq t\}}X(x_{\sigma(1)})...X(x_{\sigma(n)}) X(x_{n+\beta(1)})\dots X(x_{n+\beta(m)})$
$$\ \ \ \ \ \ \ \ \ =\int\limits_{\{0\leq t_{n+m}\leq ...\leq t_1\leq t\}}X(t_{\tau(\sigma(1))})...X(t_{\tau(\sigma(n))})X(t_{\tau(n+\beta(1))})...X(t_{\tau(n+\beta(m))})$$
so that finally, taking into account the bars of the permutations (that is the fact that $X$ is $A$ or $B$, depending only on its position in the sequence $X(t_{\tau(\sigma(1))})...X(t_{\tau(\sigma(n))})X(t_{\tau(n+\beta(1))})...X(t_{\tau(n+\beta(m))})$), we obtain 
$<{\bar\sigma}>\times <{\bar\beta}>= \sum_{\bar\gamma} <{\bar\gamma}>$,
with $st(\bar\gamma(1)\dots\bar\gamma(n))=(\bar\sigma(1),\dots,\bar\sigma(n))$ and 
$st(\bar\gamma(n+1)\dots\bar\gamma(n+m))=(\bar\beta(1),\dots,\bar\beta(n))$.
This concludes the proof.

\begin{rem}
The same proof leads to a general noncommutative
Chen formula. Let $A_1,\dots,A_n$ 
and $B_1,\dots,B_m$ be noncommutative (e.g. matrix-valued) functions and let
\begin{eqnarray*}
A_{\alpha} &=&
\int_0^t dt_1 \dots \int_0^{t_{n-1}} dt_n 
  A_1(t_{\alpha(1)})\dots A_n(t_{\alpha(n)}),\\
B_\beta &=&
\int_0^t dt_{1} \dots \int_0^{t_{m-1}} dt_{m}
 B_1(t_{\beta(1)})\dots B_m(t_{\beta(m)}),\\
(AB)_\sigma &=&
\int_0^t dt_1 \dots \int_0^{t_{n+m-1}} dt_{n+m} 
  A_1(t_{\sigma(1)})\dots A_n(t_{\sigma(n)})
 B_1(t_{\sigma(n+1)})\dots B_m(t_{\sigma(n+m)}),
\end{eqnarray*}
then $A_\alpha \times B_\beta = (AB)_{\alpha\ast\beta}$.
\end{rem}

\begin{prop}
The convolution product provides $\bf B$ with the structure of an associative (but noncommutative) algebra with a unit.
\end{prop}

The Proposition can be checked directly from the combinatorial definition of the convolution product, we refer to the original proofs \cite{ABN04,NT08}. In our setting, it also follows from the associativity of the product of iterated integrals. Notice that the general noncommutative Chen formula would relate similarly to the algebraic structures on colored permutations and wreath product group algebras introduced in \cite{NT08}.

\section{Progressions and regressions}

Modern noncommutative representation theory originates largely in the work of Solomon
\cite{solomon1976}.
From this point of view, it is natural to partition hyperoctahedral groups
into ``descent classes'', similarly to the partition of symmetric groups
into descent classes (such a partition is often referred to as a statistics on $S_n$). 

Recall that a permutation $\sigma\in S_n$ has a descent in position $i<n$ if and only if $\sigma(i)>\sigma(i+1)$. The descent set $Desc(\sigma)$ of $\sigma$ is the set of all $i<n$ such that $\sigma$ has a descent in position $i$. The partition into descent classes read: $S_n=\bigcup\limits_{I\subset [n-1]}\{\sigma ,Desc(\sigma)=I\}$. 
The \it descent algebra \rm $\mathcal D$ is the linear span of Solomon's elements $D_S^n:=\sum\limits_{\sigma\in S_n,Desc(\sigma )\subseteq S}\sigma$, where $S\subseteq [n-1]$ and $n\in \N^\ast$ (with the convention $D_\emptyset^0=1$). It is provided with a free associative algebra structure by the convolution product $\ast$ on ${\bf S}\supset\mathcal D$, see \cite[Chap.9]{reutenauer1993}.
This algebra has various natural generating families as a free
associative algebra -for instance, the family of the $D_\emptyset^n$. It is therefore also isomorphic to the algebra of noncommutative symmetric functions $\bf Sym$, from which it follows that the structure theorems for these functions can be carried back to the descent algebra -a point of view introduced and developed in \cite{G95} and a subsequent series of articles starting with \cite{KLT97}.

The corresponding descent statistics on $B_n$ is obtained by considering
the total order $\bar n<\overline{n-1}<...<\bar 1<1<...<n$. A signed permutation $\bar\sigma\in B_n$ has a descent in position $i<n$ if and only if $\bar\sigma(i)>\bar\sigma(i+1)$ \cite[Def. 3.2]{MnR95}. Descent classes are defined accordingly. The problem with this noncommutative representation theoretical statistics and with the corresponding algebraic structures is that they do not fit the needs of iterated integral computations for effective Hamiltonians, as we shall see in the forthcoming sections. Neither do the generalized descent algebras of \cite[Sect 5]{NT08}. Notice that this is not the case when symmetric groups are considered: the statistics of descent classes fits the needs of noncommutative representation theory \it as well as \rm the needs of Lie theoretical computations, as emphasized in \cite{reutenauer1993,G95}.

For this reason, we introduce another statistics on $B_n$. It seems to be new, and has surprisingly nice properties, in that it allows to generalize very naturally many algebraic properties of symmetric groups descent classes. 

We say that an element $\bar\alpha=(\alpha(1),...,\alpha(n))\in B_n$ 
has a progression in position $i$ if either: 
\begin{enumerate}
\item $|\alpha(i)|<|\alpha(i+1)|$ and $\alpha(i+1)\in {\mathbf N^\ast}$
\item $|\alpha(i)|>|\alpha(i+1)|$ and $\alpha(i+1)\in \bar{\mathbf N}^\ast$
\end{enumerate}
Else, we say that $\alpha$ has a regression in position $i$.
Here, the operation $|\ |$ is the operation of forgetting the bars, so that e.g. $|\bar 6|=6$. 
The terminology is motivated by the quantum physical idea that particles
(associated to unmarked integers) propagate forward in time, whereas
holes (associated to marked integers in our framework) propagate
backward. We refer the reader to Goldstone diagrams expansions
\cite{Goldstone} of the Gell-Mann Low eigenstate $|\Psi_{GL}>$ for
further insights into the physical motivations. Further details on these topics are contained in the following sections of this article, but we do not develop here fully the physical implications of our approach, the focus being on their mathematical background.  

We write $Reg(\alpha )$ for the set of regressions of $\alpha$. For example:
$Reg(4,\bar 3,\bar 5,6,\bar 2,1)=\{ 2,5\}$ since the sequence $(4,\bar 3,\bar 5,6,\bar 2,1)$ has only two regressions, in positions $2$ and $5$. For an arbitrary subset $S$ of $[n-1]$, we mimic now the descent statistics and write $R_S^n:=\sum\limits_{\sigma\in B_n, Reg(\sigma)=S}\sigma$. 
It is also convenient to introduce the elements $T_S^n:=\sum\limits_{\sigma\in B_n, Reg(\sigma)\subseteq S}\sigma=\sum\limits_{U\subseteq S}R_U^n$.

\begin{lem}
The elements $R_S^n$ (resp. $T_S^n$), $S\subseteq [n-1]$, form a family of linearly independent elements in the group algebra $\Q [B_n]$.
\end{lem}

The first assertion follows from the very definition of the $R_S^n$, since it is easily checked that $\{\bar\sigma\in B_n,Reg(\bar\sigma)=S\}\not=\emptyset$ for any $S\subseteq [n-1]$. The second case follows from 
the M\"obius inversion formula:
$$R_S^n=\sum\limits_{U\subseteq S}(-1)^{|S|-|U|}T_S^n,$$
where $|S|$ stands for the number of elements in $S$.

\begin{lem}\label{produit}
We have, for $S\subseteq [n-1],\ U\subseteq [m-1]$:
$$T_S^n\ast T_U^m=T_{S\cup \{n\}\cup (U+n)},$$
where $U+n=\{u+n,u\in U\}$.
\end{lem}

Indeed, by definition, for $\bar\sigma\in B_n,\ \bar\beta\in B_m$, 
with $Reg(\bar\sigma)=X\subseteq S,\ Reg(\bar\beta)=Y\subseteq U$,
$\bar\sigma\ast \bar\beta =\sum\limits_{\bar\tau} \bar\tau ,$ where
$\bar\tau$ runs over the elements of $B_{n+m}$ with 
$st(\bar\tau(1),...,\bar\tau(n))=\bar\sigma $ and 
$ st(\bar\tau(n+1),...,\bar\tau(n+m))=\bar\beta$.
In particular, for any such $\bar\tau$ and by definition of the 
standardization process:
$$Reg(\bar\tau)\subseteq X\cup\{n\}\cup (Y+n).$$

Conversely, any $\bar\tau\in B_{n+m}$ appears in the expansion of 
$st(\bar\tau(1),...,\bar\tau(n))\ast st(\bar\tau(n+1),...,\bar\tau(n+m))$ by the very definition of $\ast$
and does not appear in the expansion of any other product 
$\bar\sigma\ast \bar\beta$ with 
$Reg(\bar\sigma)=Reg(st(\bar\tau(1),...,\bar\tau(n)))$, 
$Reg(\bar\beta)=Reg(st(\bar\tau(n+1),...,\bar\tau(n+m))$, from which the 
lemma follows.

\begin{cor}
For $S,U$ as above, with the notation of the previous sections:
$$<{T_S^n}>\times <{T_U^m}>=<{T_{S\cup\{n\}\cup (U+n)}^{n+m}}>$$
so that:
$$<{T_\emptyset^{n_1}}>\times ...\times <{T_\emptyset^{n_k}}>= 
<{T_{\{n_1, ...,n_1+...+n_{k-1}\}}^{n_1+...+n_k}}> .$$
\end{cor}

\begin{thm}\label{Reg}
The linear span $\mathcal R$ of the elements $T_S^n$ (equivalently, of
the $R_S^n$), $n\in\N , \ S\subseteq [n-1]$, is closed under the convolution
product in $\bf B$. This algebra, referred to from now on as the (hyperoctahedral) Regression algebra, is isomorphic to the descent algebra $\mathcal D$ and to the algebra of noncommutative symmetric functions $\bf Sym$.
\end{thm}

The second part of the Theorem follows from the product rule in $\cal D$, that reads:
$$D_S^n\ast D_U^m=D_{S\cup\{n\}\cup (U+n)}^{n+m}.$$
The proof for this last identity can be obtained similarly to the one in Lemma~\ref{produit} -see also \cite{reutenauer1993}.

Now we study in more detail the elements $R_\emptyset^n$
that will play an important role in the following.
The lowest order $R_\emptyset^n$ are
\begin{eqnarray*}
R_\emptyset^1 &=& (1) + (\bar1),\\
R_\emptyset^2 &=& (1,2) +  (\bar1,2) + (2,\bar1) + (\bar2,\bar1),\\
R_\emptyset^3 &=& (1,2,3) +  (\bar1,2,3) + 
   (1,3,\bar2)+(\bar1,3,\bar2)+(2,\bar1,3)+(\bar2,\bar1,3)
  +(2,3,\bar1)+(\bar2,3,\bar1)
  \\&&+(3,\bar1,2)+(\bar3,\bar1,2)+(3,\bar2,\bar1)+(\bar3,\bar2,\bar1).
\end{eqnarray*}
We first observe that, if $\bar\sigma \in B_n$ is a term of $
R^n_\emptyset$ (and therefore has no regression), then the barred integers of $\bar\sigma$
are entirely determined by the permutation
$\sigma=(|\bar\sigma(1)|,\dots,|\bar\sigma(n)|)$,
except for $\bar\sigma(1)$. Indeed, by definition of a progression,
$\bar\sigma(i+1)\in {\mathbf N}^\ast$ if 
$\sigma(i)<\sigma(i+1)$ and
$\bar\sigma(i+1)\in \bar{\mathbf N}^\ast$ if 
$\sigma(i)>\sigma(i+1)$.
In other words, 
$\bar\sigma(i+1)\in \bar{\mathbf N}^\ast$ iff $\sigma$ has
a descent at $i$. 
The integer $\bar\sigma(1)$ is not determined by $\sigma$ and can be
barred or not.
Therefore, the number of terms of $R_\emptyset^n$ is $2\cdot n!$.

\section{A Picard-type hyperoctahedral expansion}
When it comes to expand $\Psi_{GL}$ or $U_{GL}$, as introduced in Section 2, the classical strategy introduced by Goldstone (at least for nondegenerate states, that is for $\Psi_{GL}$ \cite{Goldstone}) consists in appealing to the hole/particle duality of quantum physics. Goldstone's theory was generalized to degenerate states by Michels and Suttorp \cite{ms}, but this part of the theory has remained largely in infancy and relies on shaky mathematical grounds. The purpose of this section is to show that hyperoctahedral groups provide a convenient way to derive and study such expansions, so as to build the foundations of a group-theoretic approach to the perturbative computation of the ground states of physical systems, with a particular view toward the degenerate case.

To sum up, we want to compute $U_{GL}(\epsilon)=U_\epsilon(0)P (PU_\epsilon(0)P)^{-1}$. 
Let us write $H_i$ for $-i\Hint$ and $A(t):=(1-P)H_i(t)$, $B(t):=-PH_i(t)$ 
(notice the $-1$ sign in the definition of $B$). From the Picard expansion, 
we have:
$$U_\epsilon(0)=1+\int\limits_{-\infty}^{0}H_i(x)dx+\int\limits_{-\infty}^{0}\int\limits_{-\infty}^{t_1}H_i(t_1)H_i(t_2)dt_1dt_2+...+\int\limits_{\Delta_n^{[-\infty ,0]} } H_i(t_1)...H_i(t_n)+...$$
In this section (following a suggestion by the referee whose remarks helped us to simplify notably the presentation of the following computations -we take the opportunity to thank him or her warmly), 
we introduce a new encoding of iterated integrals involving $A$, $B$ and $H_i$.

The notation is best explained through an example: $$\int\limits_{-\infty}^{0}\int\limits_{-\infty}^{t_1}\int\limits_{-\infty}^{t_2}\int\limits_{-\infty}^{t_3}\int\limits_{-\infty}^{t_4}A(t_3)H_i(t_1)B(t_2)B(t_5)H_i(t_4)dt_1dt_2dt_3dt_4dt_5=:[\hat 2,\bar 3,1,\hat 5,\bar 4].$$
Concretely, in an arbitrary iterated integral involving $A$, $B$ and $H_i$, we look recursively at the positions $i_1,...,i_n$ of $t_1$,...,$t_n$ in the integrand and decorate $i_j$ with a bar (resp. a hat, resp. no decoration) if the corresponding operator is $B$ (resp. $H_i$, resp. $A$). In our example, $t_1$ (resp. $t_2$...) is in position 2 (resp. 3...) in the product $A(t_3)H_i(t_1)B(t_2)B(t_5)H_i(t_4)$ and appears as a parameter for $H_i$ (resp. $B$), so that we map 1 to $\hat 2$, 2 to $\bar 3$, and so on.
The so-obtained sequence of decorated integers is written inside brackets to avoid confusion with our previous notation. Notice that, if $\bar\sigma\in B_n$,
$<{\bar\sigma}>=[\bar\sigma^{-1}(1),...,\bar\sigma^{-1}(n)]$ (where we use the definition of Section~\ref{sect3} for $<{\bar\sigma}>$).
We will also use some self-explaining multilinear extensions of this notation, so that, for example: $[3,(\bar 1,\hat 2)+(2,\hat 1)]=[3,\bar 1,\hat 2]+[3,2,\hat 1]$, and so on. Notice in particular that the identity $H_i=A-B$ translates formally into $\hat k=k-\bar k$ so that, for example,
$[\hat 1,\bar 2]=[1,\bar 2]-[\bar 1,\bar 2]$.

\begin{thm}\label{thmfond}
The effective adiabatic evolution operator $U_{GL}$ has the hyperoctahedral Picard-type expansion:
$$U_{GL}=\lim\limits_{\epsilon \to 0}P+(1-P)(\sum\limits_{n\in\mathbf N^\ast}<{R_\emptyset^n}>)P$$
\end{thm}

Indeed, let us expand $[\hat 1,...,\hat n]=\int\limits_{\Delta_n^{[-\infty ,0]} } H_i(t_1)...H_i(t_n)$ with the $A$ and $B$ operators. 
In order to do so, we introduce still another notation: for $\bar\sigma\in B_{k}$, $k<n$, we set:
$$<\bar\sigma ; n-k>=\int\limits_{\Delta_k^{[-\infty ,0]}\times
\Delta_{n-k}^{[-\infty
,t_{\sigma(k)}]}}X(t_{\sigma(1)})...X(t_{\sigma(k)})H_{i}(t_{k+1})...H_{i}(t_n)$$
where $\Delta_k^{[-\infty ,0]}\times \Delta_{n-k}^{[-\infty ,t_{\sigma(k)}]}$ is a shortcut for:
$$\{(t_1,...,t_n)|-\infty\leq t_k\leq ...\leq t_1\leq 0,\ -\infty \leq t_{n}\leq ...\leq t_{k+1}\leq t_{\sigma(k)}\};$$
where $\sigma$ stands, as usual, for the image of $\bar\sigma$ in $S_k$ (obtained by forgetting the decorations), and where $X(t_{\sigma(i)})=A(t_{\sigma(i)})$ if $\sigma(i)=\bar\sigma(i)$ and $B(t_{\sigma(i)})$ else. For example,
$$<(2\bar1 3);2>=\int\limits_{\Delta_3^{[-\infty
,0]}\times\Delta_2^{[-\infty ,t_3]} }A(t_2)B(t_1)A(t_3)H_{i}(t_4)H_{i}(t_5),$$
$$<(2\bar3 1);2>=\int\limits_{-\infty\leq t_3\leq t_2\leq t_1\leq 0,\
-\infty\leq t_5\leq t_4\leq t_1}A(t_2)B(t_3)A(t_1)H_{i}(t_4)H_{i}(t_5).$$
The more general symbols $<X;n-k>$ are defined, as usual, by extending linearly these conventions to arbitrary elements $X\in \Q[B_{k}]$, $k<n$.

\begin{lem}\label{lem51}
The symbol $<\bar\sigma ; n-k>$ can be expanded as:
$$<\bar\sigma ;
n-k>=[\bar\sigma^{-1}(1),...,\bar\sigma^{-1}(i),(\bar\sigma^{-1}(i+1),...,\bar\sigma^{-1}(k))\shuffle
(\widehat{k+1},...,\hat{n})],$$
where $\bar\sigma\in B_k$ and $i:=\sigma(k)$.
\end{lem}

Indeed:
$$\{(t_1,...,t_n)|-\infty\leq t_k\leq ...\leq t_1\leq 0,\ -\infty \leq t_{n}\leq ...\leq t_{k+1}\leq t_{\sigma(k)}\}=$$
$$\{(t_1,...,t_n)|-\infty\leq t_k\leq ...\leq t_{\sigma(k)},\ -\infty \leq t_{n}\leq ...\leq t_{k+1}\leq t_{\sigma(k)}, t_{\sigma(k)}\leq ...\leq t_1\leq 0\}$$
$$=\{(t_1,t_2,...,t_n)|-\infty\leq t_{\sigma(k)}\leq ...\leq t_1\leq 0,(t_{\sigma(k)+1},...,t_n)\in $$ $$\{(u_1,...,u_{k-\sigma(k)})|-\infty\leq u_{\sigma(k)}\leq ...\leq u_1\leq t_{\sigma(k)}\}\times \{(v_1,...,v_{n-k})|-\infty\leq v_{n-k}\leq ...\leq v_1\leq t_{\sigma(k)}\}
\},$$
where $\times$ stands for the cartesian product. Since the Cartesian product of simplices is reflected in the shuffle product (see e.g. the proof of Prop.~\ref{HsHb}),
the Lemma follows.

We then have:
\begin{eqnarray*}
[\hat 1,...\hat n]&=&[1,\hat 2,...\hat n]-[\bar 1,\hat 2,...,\hat n]=-[\bar 1,\hat 2,...,\hat n]+[1,2,\hat 3,...\hat n]-[1,\bar 2,\hat 3,...,\hat n]\\
&=&P[\hat 1,...\hat n]+[1,2,\hat 3,...\hat n]-[1,\bar 2,\hat 3,...,\hat n]\\
&=&P[\hat 1,...\hat n]+[1,2,\hat 3,...\hat n]-[1,\bar 2,\hat 3,...,\hat n]+(-[\bar2,(1)\shuffle (\hat 3,...,\hat n)]+[\bar2,(1)\shuffle (\hat 3,...,\hat n)]).
\end{eqnarray*}

Now, from the recursive definition of the shuffle product:
$$[1,\bar 2,\hat 3,...,\hat n]+[\bar2,(1)\shuffle (\hat 3,...,\hat
n)]=[(1)\shuffle (\bar2,\hat 3,...,\hat n)]$$
$$=[\int\limits_{-\infty}^0A(t)dt]\int\limits_{\Delta_{n-1}^{[-\infty ,0]} }B(t_1)H_i(t_2)...H_i(t_{n-1})$$
$$=-(1-P)<{R_\emptyset^1}>P[\hat 1,...,\widehat{n-1}].$$
Here, we have used that $P$ is a projection, so that $(1-P)B(t)=-(1-P)PH_i(t)=0$, to rewrite $$[\int\limits_{-\infty}^0A(t)dt]=(1-P)[\int\limits_{-\infty}^0(A(t)+B(t))dt]=(1-P)<{R_\emptyset^1}>.$$

We get finally, since $<(1, 2);n-2>=[1, 2, \hat 3,...,\hat n]$ and (according to the Lemma~\ref{lem51})
$<(2,\bar1);n-2>=[\bar2,(1)\shuffle (\hat 3,...,\hat n)]$:
$$
[\hat 1,...,\hat n]=P[\hat 1,...,\hat n]+(1-P)<{R_\emptyset^1}>P[\hat 1,...,\widehat{n-1}]+(1-P)<R_\emptyset^2;n-2>,$$
where the last identity follows, once again, from $(1-P)B(t)=0$ (we won't comment any more on this rewriting trick from now on).

The proof of the Theorem can be obtained along these principles by recursion. Let us indeed assume for a while that:
$$<{R_\emptyset^k;n-k}>=<{R_\emptyset^k}>P[\hat 1,...,\widehat{n-k}]+<{R_\emptyset^{k+1};n-k-1}>.$$
Then we get, by induction:
$$[\hat 1,...,\hat n]=P[\hat 1,...,\hat n]+(1-P)<{R_\emptyset^1}>P[\hat 1,...,\widehat{n-1}]+$$
$$(1-P)<{R_\emptyset^2}>P[\hat 1,...,\widehat{ n-2}]+...+(1-P)<{R_\emptyset^{n-1}}>P[\hat 1]+(1-P)<{R_\emptyset^n}>.$$
Since $U_\epsilon(0)=\sum_n[\hat 1,...,\hat n]$, this implies
\begin{eqnarray*}
U_\epsilon(0)&=& P U_\epsilon(0) + (1-P) \sum_{n=1}^\infty  <{R_\emptyset^n}>( P(U_\epsilon(0)-1)+1),
\end{eqnarray*}
or, since $P^2=P$:
\begin{eqnarray*}
U_\epsilon(0) P &=& \Big( P + (1-P) \sum_{n=1}^\infty  <{R_\emptyset^n}> P\Big) PU_\epsilon(0)P,
\end{eqnarray*}
and the Theorem follows.

So, let us check that the formula for $<{R_\emptyset^k;n-k}>$ holds. 
Let us consider an arbitrary element $\bar\sigma \in B_k$ with 
$Reg(\bar\sigma)=\emptyset$. Then, with the notation of Lemma~\ref{lem51}:
$$
<{\bar\sigma;n-k}>
=[\bar\sigma^{-1}(1),...,\bar\sigma^{-1}(i),(\bar\sigma^{-1}(i+1),...,\bar\sigma^{-1}(k))\shuffle
(k+1,\widehat{k+2},...,\hat n)]$$
$$-[\bar\sigma^{-1}(1),...,\bar\sigma^{-1}(i),(\bar\sigma^{-1}(i+1),...,\bar\sigma^{-1}(k))\shuffle
(\overline{k+1},\widehat{k+2},...,\hat n)]
$$
Let us denote the first term by $T_1$ and the second by $T_2$, so that $<{\bar\sigma;n-k}>=T_1-T_2$.
To calculate $T_1$, let us use another (equivalent, the equivalence follows from the recursive definition of the shuffle product and is left to the reader) recursive definition of the shuffle product, namely:
$$a_1a_2...a_k\shuffle bb_2...b_l=\sum\limits_{i=0}^ka_1...a_ib(a_{i+1}...a_k\shuffle b_2...b_l).$$
We get:
$$T_1=\sum\limits_{j=i}^k[\bar\sigma^{-1}(1),...,\bar\sigma^{-1}(j),k+1,(\bar\sigma^{-1}(j+1),...,\bar\sigma^{-1}(k))\shuffle
(\widehat{k+2},...,\hat n)]$$
$$=\sum\limits_{j=i}^k<\bar\beta_j;n-k-1>$$
where $\bar\beta_j:=(\bar\sigma^{-1}(1),...,\bar\sigma^{-1}(j),k+1,\bar\sigma^{-1}(j+1),...,\bar\sigma^{-1}(k))^{-1}$.
We notice then that, for $l<k$,
$$\beta_j(l)<\beta_j(l+1)\Longleftrightarrow\sigma(l)<\sigma(l+1).$$
In particular, $\bar\beta_j$ has no regression in position less than $k$. Now, $\beta_j(k+1)=\bar\beta_j(k+1)=j\geq \beta_j(k)=i$, which implies that $\bar\beta_j$ has no regression in position $k$. Finally, $\bar\beta_j$ has no regression.

Let us enumerate the number of (necessarily distinct) signed permutations $\bar\beta_j$ obtained in that way.
There are $2(k-1)!$ elements $\bar\sigma$ of $R^k_\emptyset$ with
a given value $j$ of $\sigma(k)$, where $j$ runs from 1 to $k$. 
For a given $\bar\sigma$ with $\sigma(k)=j$, $T_1$ provides $k-j+1$ elements
of $R^{k+1}_\emptyset$. Thus, the expansion of $T_1$ provides $(k+1)!$ different elements
of $R^{k+1}_\emptyset$ when $\bar\sigma$ runs over  $R^k_\emptyset$.

The term $T_2$ can be computed similarly.
Using the same recursive formula for the shuffle product as above, we get:
$$T_2=\sum\limits_{j=i}^k[\bar\sigma^{-1}(1),...,\bar\sigma^{-1}(j),\overline{k+1},(\bar\sigma^{-1}(j+1),...,\bar\sigma^{-1}(k))\shuffle
(\widehat{k+2},...,\hat n)]$$
$$=[(\bar\sigma^{-1}(1),...,\bar\sigma^{-1}(k))\shuffle (\overline{k+1},...,\bar n)]$$
$$-\sum\limits_{j<i}[(\bar\sigma^{-1}(1),...,\bar\sigma^{-1}(j),\overline{k+1},(\bar\sigma^{-1}(j+1),...,\bar\sigma^{-1}(k))\shuffle
(\widehat{k+2},...,\hat n)]$$
In the first term in this expansion of $T_2$, we recognize $-<\bar\sigma>P[\hat 1,...,\widehat{n-k}]$, so that these terms sum up to
$-<R_\emptyset^k>P[\hat 1,...,\widehat{n-k}]$ when $\bar\sigma$ runs over elements in $B_k$ without regressions. In the second term (the sum over $j<i$), the same reasoning as for $T_1$ shows that each term of the expansion is of the form $<\bar\beta_j;n-k-1>$, where $\bar\beta_j$ has no regression. Again, when $\bar\sigma$ runs over elements in $B_k$ without regressions, this provides $(k+1)!$ such elements in the expansion. These terms are pairwise distinct and pairwise distinct from the elements showing up in the expansion of $T_1$, from which we conclude:
$$T_1-T_2=<R_\emptyset^k>P[\hat 1,...,\widehat{n-k}]+<R_\emptyset^{k+1};n-k+1>.$$
The Theorem follows.

\section{A Magnus expansion for the evolution operator}

In the classical case, that is when the solution $X(t)$ of a first order linear differential equation is obtained from its Picard series expansion, the resulting approximating series converges relatively slowly to the solution. This problem --let us call it the Magnus problem-- is solved by reorganizing the series expansion, often by looking for an exponential expansion $X(t)=\exp{\Omega (t)}$ of the solution, known as its Magnus expansion. Many numerical techniques have been developed along this idea that go much beyond the formal-algebraic problem of deriving a formal expression for $\Omega(t)$. However, deriving such an expression is a decisive step towards the understanding of the behavior of $\Omega(t)$. This problem was solved, in the classical case, by Bialynicki-Birula, Mielnik and Pleba\'nski \cite{BMP69,MP70} who obtained a formula  for $\Omega (t)$ in terms of Solomon's elements $D_S^n$.

The purpose of the present section is to solve the Magnus problem for
the analysis of solutions in time-dependent perturbation theory. This
provides the general term of time-dependent 
coupled-cluster theory \cite{Schonhammer}. Our previous results pave the way toward the solution of the problem. Namely, as it appears from Thm~\ref{thmfond}, the natural object to look at is not so much the effective Hamiltonian
$${\cal H}=\lim\limits_{\epsilon\rightarrow 0} P_0 H U_{GL}(\epsilon)$$
or the effective adiabatic evolution operator $U_{GL}$, than the Picard-type series $$Pic:=\sum\limits_{n\in \N}<{R_\emptyset^n}>.$$
Notice that we define $Pic$ as the sum of the $<{R_\emptyset^n}>$ over all the integers (and not over $\N^\ast$) in order to have the identity operator $1=H_{R_\emptyset^0}$ as the first term of the series. Of course, we have:
$$U_{GL}(\epsilon)=P+(1-P)(\sum\limits_{n\in\N^\ast}<{R_\emptyset^n}>)P=P+(1-P)(\sum\limits_{n\in\N}<{R_\emptyset^n}>)P=P+(1-P)\ Pic \ P$$
In other terms, we are interested in the expansion:
$$U_{GL}(\epsilon)=P+(1-P)\exp(\Omega_{\epsilon})P,$$
where 
$$\Omega_{\epsilon}=\log(\sum\limits_{n\in\N}<{R_\emptyset^n}>)=<{\log(\sum\limits_{n\in\N}R_\emptyset^n)}>.$$
Since $R_\emptyset^{n_1}\ast ...\ast
R_\emptyset^{n_k}=R_{\{n_1,...,n_1+...+n_{k-1}\}}^{n_1+...+n_k},$ a first expression of $\Omega_R=\log{\sum\limits_{n\in\N}R_\emptyset^n}$ follows:
$$\Omega_R=\sum\limits_{n\in\N^\ast}\sum\limits_{S\subseteq [n-1]}\frac{(-1)^{|S|}}{|S|+1}R_S^n,$$
where one can recognize the hyperoctahedral analogue of Solomon's Eulerian idempotent \cite[Chap.3, Lem.3.14]{reutenauer1993}:
$$sol_n=\sum\limits_{S\subseteq [n-1]}\frac{(-1)^{|S|}}{|S|+1}D_S^n.$$
The analogy is not merely formal and follows from the isomorphism of Thm~\ref{Reg} together with the existence of a logarithmic expansion of $sol_n$, which is actually best understood from an Hopf algebraic point of view, see \cite{pat92,reutenauer1993,pat93,pat94}:
$$\sum\limits_{n\in\N^\ast}sol^n=\log(\sum_{n\in\N}D_\emptyset^n).$$

As a corollary of Thm~\ref{Reg}, we also get the expansion of $\Omega_R$ in the canonical basis of $\bigoplus\limits_{n\in\N^\ast}\Q[B_n]$:
\begin{prop}
We have:
$$\Omega_R=\sum\limits_{n\in\N^\ast}\sum\limits_{S\subseteq[n-1]}\frac{(-1)^{|S|}}{n}{{n-1}\choose{|S|}}^{-1}T_S^n$$
$$=\sum\limits_{n\in\N^\ast}\sum\limits_{S\subseteq[n-1]}\sum\limits_{\bar\sigma\in B_n,Reg(\sigma )=S}\frac{(-1)^{|S|}}{n}{{n-1}\choose{|S|}}^{-1}\bar\sigma$$
\end{prop}

The Proposition follows from the analogous expansion for $sol_n$ \cite{reutenauer1993}, together with the algebra isomorphism Thm~\ref{Reg}:
$$sol_n=\sum\limits_{n\in\N^\ast}\sum\limits_{S\subseteq[n-1]}\sum\limits_{\sigma\in S_n,Desc(\sigma )=S}\frac{(-1)^{|S|}}{n}{{n-1}\choose{|S|}}^{-1}\sigma .$$

\begin{cor}
The hyperoctahedral Magnus expansion of the effective Hamiltonian $\mathcal H$ reads, when truncated at the third order:
$${\mathcal H}=\lim\limits_{\epsilon\rightarrow 0}P H_I(P+(1-P)\exp (H_{(1)}+H_{(\bar 1)}+\frac{1}{2}[H_{(12)}+H_{(\bar 12)}+H_{(2\bar 1)}+H_{(\bar 2\bar 1)}-H_{(1\bar 2)}-H_{(\bar 1\bar 2)}-H_{(21)}-H_{(\bar 21)}]+$$
$$\frac{1}{3}[H_{(123)}+H_{(\bar 123)}+H_{(13\bar 2)}+H_{(\bar 13\bar 2)}+H_{(2\bar 13)}+H_{(\bar 2\bar 13)}
+H_{(23\bar 1)}+H_{(\bar 23\bar 1)}+H_{(3\bar 2\bar 1)}+H_{(\bar 3\bar 2\bar 1)}+H_{(3\bar 12)}+H_{(\bar 3\bar 12)}$$
$$+H_{(321)}+H_{(\bar 321)}+H_{(2\bar 31)}+H_{(\bar 2\bar 31)}+H_{(1\bar 2\bar 3)}+H_{(\bar 1\bar 2\bar 3)}+H_{(1\bar 32)}+H_{(\bar 1\bar 32)}+H_{(21\bar 3)}+H_{(\bar 21\bar 3)}+H_{(31\bar 2)}+H_{(\bar 31\bar 2)}]$$
$$-\frac{1}{6}[H_{(132)}+H_{(\bar 132)}+H_{(231)}+H_{(\bar 231)}+H_{(213)}+H_{(\bar 213)}+H_{(312)}+H_{(\bar 312)}+H_{(1\bar 3\bar 2)}+H_{(\bar 1\bar 3\bar 2)}+H_{(1\bar 23)}+H_{(\bar 1\bar 23)}+H_{(2\bar 1\bar 3)}$$
$$+H_{(\bar 2\bar 1\bar 3)}+H_{(3\bar 1\bar 2)}+H_{(\bar 3\bar 1\bar 2)}+H_{(2\bar 3\bar 1)}+H_{(\bar 2\bar 3\bar 1)}+H_{(3\bar 21)}+H_{(\bar 3\bar 21)}+H_{(12\bar 3)}+H_{(\bar 12\bar 3)}+H_{(32\bar 1)}+H_{(\bar 32\bar 1)}  ]
)P).$$
\end{cor}


\begin{thebibliography}{99}
\bibitem{ABN04}
    M. Aguiar, N. Bergeron and K. Nyman, 
    The peak algebra and the descent algebra of type $B$ and $D$,
    \textit{Trans. of the AMS}
    356, (2004), 2781-2824.
    
\bibitem{ANT08}
    M. Aguiar, J.-C. Novelli and J.-Y. Thibon, 
    Unital versions of the higher peak algebras,
    arXiv:0810.463v1.    
    
\bibitem{BMP69}
    I. Bialynicki-Birula, B. Mielnik and J. Pleba\'nski,
    Explicit solution of the continuous Baker-Campbell-Hausdorff problem,
    \textit{Annals of Physics}
    51, (1969), 187--200.

\bibitem{Bloch}
    C. Bloch,
    Sur la th{\'e}orie des perturbations des {\'e}tats li{\'e}s,
    \textit{Nuclear Physics}
    6, (1958), 329--347.

\bibitem{BH06}
    C. Bonnaf\'e and C. Hohlweg,
    Generalized descent algebra and construction of irreducible characters of hyperoctahedral groups,
    \textit{Ann. Inst. Fourier}
    56, (2006), 131--181.    
    
\bibitem{bulaevskii}
    L.N. Bulaevskii,
    \textit{Soviet Phys. JETP}
    24 (1967) 154.  

\bibitem{chen1973}
    K.T. Chen,
    Iterated integrals of differential forms and loop space homology,
    Annals of Math. 97 (1973), 217-246.
     
\bibitem{DHT02}
    G. Duchamp, F. Hivert, and J.-Y. Thibon, 
    Noncommutative symmetric functions VI: free quasi-symmetric functions and related algebras, 
    \textit{Internat. J. Alg. Comput. }
    12 (2002), 671–717.
    
\bibitem{G95}
   I.M. Gelfand, D. Krob, A. Lascoux, B. Leclerc, V. S. Retakh, and J.-Y. Thibon,
   Noncommutative symmetric functions, 
   \textit{Adv. in Math.}
   112 (1995), 218–348.    
    
\bibitem{GellMann}
M.~Gell-Mann and F.~Low.
\newblock Bound states in quantum field theory.
\newblock {\em Phys. Rev.}, 84:350--4, 1951.

\bibitem{Goldstone} 
    J. Goldstone,
    Derivation of the Brueckner many-body theory.
    \textit{Lectures on the many-body problem}
    Proc. Roy. Soc. London,
    A239 (1957) 267--79.
    
\bibitem{KLT97}
    D. Krob, B. Leclerc, and J.-Y. Thibon, 
    Noncommutative symmetric functions II: Transformations of alphabets, 
    \textit{Internal J. Alg. Comput.}
    7 (1997), 181–264.

\bibitem{MR95}    
    C. Malvenuto and Ch. Reutenauer,
    Duality between quasi-symmetric functions and Solomon's descent algebra
    \textit{J. Algebra}
    177, (1995), 967--982.
    
\bibitem{MnR95} 
    R. Mantaci and C. Reutenauer,
    A generalization of Solomon's algebra for hyperoctahedral groups and other wreath products,
    \textit{Comm. Algebra}
    23, (1995), 27--56.
        
\bibitem{ms}
    M.A.J. Michels and L.G. Suttorp,
    Diagrammatic analysis of adiabatic and time-independent perturbation theory for degenerate energy levels.
    \textit{Physica A}
    58, (1978), 385-388.    
    
    
\bibitem{MP70}
    B. Mielnik and J. Pleba\'nski,
    Combinatorial approach to Baker-Campbell-Hausdorff exponents,
    \textit{Ann. Inst. Henri Poincar\'e, Section A}
    Vol.~XII, (1970), 215--254.

\bibitem{morita} T. Morita,
    Progr. Theoret. Phys. 
    29 (1963) 351.    
        
\bibitem{Nenciu}
    G.~Nenciu and G.~Rasche.
    \newblock Adiabatic theorem and {G}ell-{M}ann-{L}ow formula.
    \newblock {\em Helv. Phys. Acta}, 62:372--88, 1989.
        
\bibitem{NT08}
    J.-C. Novelli and J.-Y. Thibon, 
    Free quasi-symmetric functions and descent algebras for wreath products and noncommutative multisymmetric functions, 
    arXiv:0806.3682v1.      
        
\bibitem{pat92}
    F.~Patras,
    Homoth\'eties simpliciales,
    Ph.D. Thesis, Univ. Paris 7, Jan. 1992.         
        
\bibitem{pat93}
    F.~Patras,
    La d\'ecomposition en poids des alg\`ebres de Hopf,
    \textit{Ann.~Inst.~Fourier} 43 (1993) 1067--1087.

\bibitem{pat94}
    F.~Patras,
    L'alg\`ebre des descentes d'une big\`ebre gradu\'ee,
    \textit{J. Algebra} 170 (1994) 547--566.        

\bibitem{patreu02}
    F.~Patras and C.~Reutenauer,
    On Dynkin and Klyachko idempotents in graded bialgebras,
    \textit{Adv. Appl. Math.} 28 (2002) 560--579.

\bibitem{reutenauer1986}
    C. Reutenauer,
    Theorem of Poincar\'e-Birkhoff-Witt, logarithm and representations of the symmetric group whose orders are the Stirling numbers. 
    \textit{Combinatoire enum\'erative}, Proceedings, Montr\'eal (1985), (ed. G. Labelle and P. Leroux). Lecture Notes in Mathematics, 267--284, Springer, Berlin.
    
\bibitem{reutenauer1993}
    C.~Reutenauer,
    Free Lie algebras,
    Oxford University Press, Oxford, 1993.

\bibitem{Schonhammer}
    K. Sch{\"o}nhammer and O. Gunnarsson,
    Time-dependent approach to the calculation of spectra functions,
    \textit{Phys. Rev. B}
    18, (1978), 6606--6614.
   
\bibitem{sch58}
   M. P. Sch\"utzenberger,
   {\textsl{Sur une propri\'et\'e combinatoire des alg\`ebres de Lie libres pouvant \^etre utilis\'ee dans un probl\`eme de math\'ematiques appliqu\'ees}}
   S\'eminaire Dubreil--Jacotin Pisot (Alg\`ebre et th\'eorie des nombres), Paris, Ann\'ee 1958/59.    

\bibitem{solomon1968}
    L.~Solomon,
    On the Poincar\'e-Birkhoff-Witt theorem,
    \textit{J. Combinatorial Theory} 4 (1968) 363--375.
    
\bibitem{solomon1976}
    L. Solomon, 
    A Mackey formula in the group algebra of a finite Coxeter group, 
    \textit{J. Algebra} 41 (1976) 255–268.
   
\end{thebibliography}
\end{document}